\pdfoutput=1
 
\newcommand*{\ATLASLATEXPATH}{}
 
\documentclass[PAPER, cernpreprint, atlasdraft=false, texlive=2016, UKenglish]{\ATLASLATEXPATH atlasdoc}
 
\usepackage{\ATLASLATEXPATH atlaspackage}
 
\usepackage{\ATLASLATEXPATH atlasbiblatex}
 
\usepackage{\ATLASLATEXPATH atlasphysics}
 
\graphicspath{{logos/}{figures/}}
 
\usepackage{tabularx}
\newcolumntype{L}[1]{>{\raggedright\arraybackslash}p{#1}}
\newcolumntype{C}[1]{>{\centering\arraybackslash}p{#1}}
\newcolumntype{R}[1]{>{\raggedleft\arraybackslash}p{#1}}
 
\AtlasTitle{Search for flavour-changing neutral currents in processes with one top quark and a photon using 81~fb$^{-1}$ of $pp$ collisions at $\sqrt{s} = 13$~\TeV\ with the ATLAS experiment}
 
\AtlasAbstract{
A search for flavour-changing neutral current (FCNC) events via the coupling of a top quark, a photon, and an up or charm quark is presented using 81~fb$^{-1}$ of proton--proton collision data taken at a centre-of-mass energy of 13~\TeV\ with the ATLAS detector at the LHC. Events with a photon, an electron or muon, a $b$-tagged jet, and missing transverse momentum are selected. A neural network based on kinematic variables differentiates between events from signal and background processes. The data are consistent with the background-only hypothesis, and limits are set on the strength of the $tq\gamma$ coupling in an effective field theory. These are also interpreted as 95\% CL upper limits on the cross section for FCNC $t\gamma$ production via a left-handed (right-handed) $tu\gamma$ coupling of 36~fb (78~fb) and on the branching ratio for $t\rightarrow \gamma u$ of $2.8\times 10^{-5}$ ($6.1\times 10^{-5}$). In addition, they are interpreted as 95\% CL upper limits on the cross section for FCNC $t\gamma$ production via a left-handed (right-handed) $tc\gamma$ coupling of 40~fb (33~fb) and on the branching ratio for $t\rightarrow \gamma c$ of $22\times 10^{-5}$ ($18\times 10^{-5}$).
}
 
\PreprintIdNumber{CERN-EP-2019-155}
 
\AtlasDate{\today}
 
\AtlasJournalRef{Phys. Lett. B 800 (2019) 135082}
\AtlasDOI{10.1016/j.physletb.2019.135082}
\hypersetup{pdftitle={ATLAS document},pdfauthor={The ATLAS Collaboration}}
 
\newcommand{\egfake}{\ensuremath{f_{e\rightarrow\gamma}}}
 
\newcommand{\AtlasCoordFootnote}{
ATLAS uses a right-handed coordinate system with its origin at the nominal interaction point (IP)
in the centre of the detector and the \(z\)-axis along the beam pipe.
The \(x\)-axis points from the IP to the centre of the LHC ring,
and the \(y\)-axis points upwards.
Cylindrical coordinates \((r,\phi)\) are used in the transverse plane,
\(\phi\) being the azimuthal angle around the \(z\)-axis.
The pseudorapidity is defined in terms of the polar angle \(\theta\) as \(\eta = -\ln \tan(\theta/2)\).
Angular distance is measured in units of \(\Delta R \equiv \sqrt{(\Delta\eta)^{2} + (\Delta\phi)^{2}}\).}
 
\begin{document}
 
\maketitle
 
\section{Introduction}
Flavour-changing neutral currents (FCNCs) are forbidden at tree level in the Standard Model (SM) and strongly suppressed at higher orders via the GIM mechanism~\cite{Glashow:1970gm}. Several extensions to the SM predict processes involving FCNCs. In particular, some of these models predict the branching ratios of top-quark decays via FCNC to be significantly larger~\cite{AguilarSaavedra:2004wm} than those predicted by the SM, which are of the order of $10^{-14}$~\cite{AguilarSaavedra:2004wm}. Examples are R-parity-violating supersymmetric models~\cite{Yang:1997dk} and models with two Higgs doublets~\cite{Atwood:1996vj}. Such models would allow the production of top quarks via FCNCs at a measurable rate.
 
\begin{figure}[ht!]
\begin{center}
\begin{tabular}{cc}
\includegraphics[width=0.49\columnwidth]{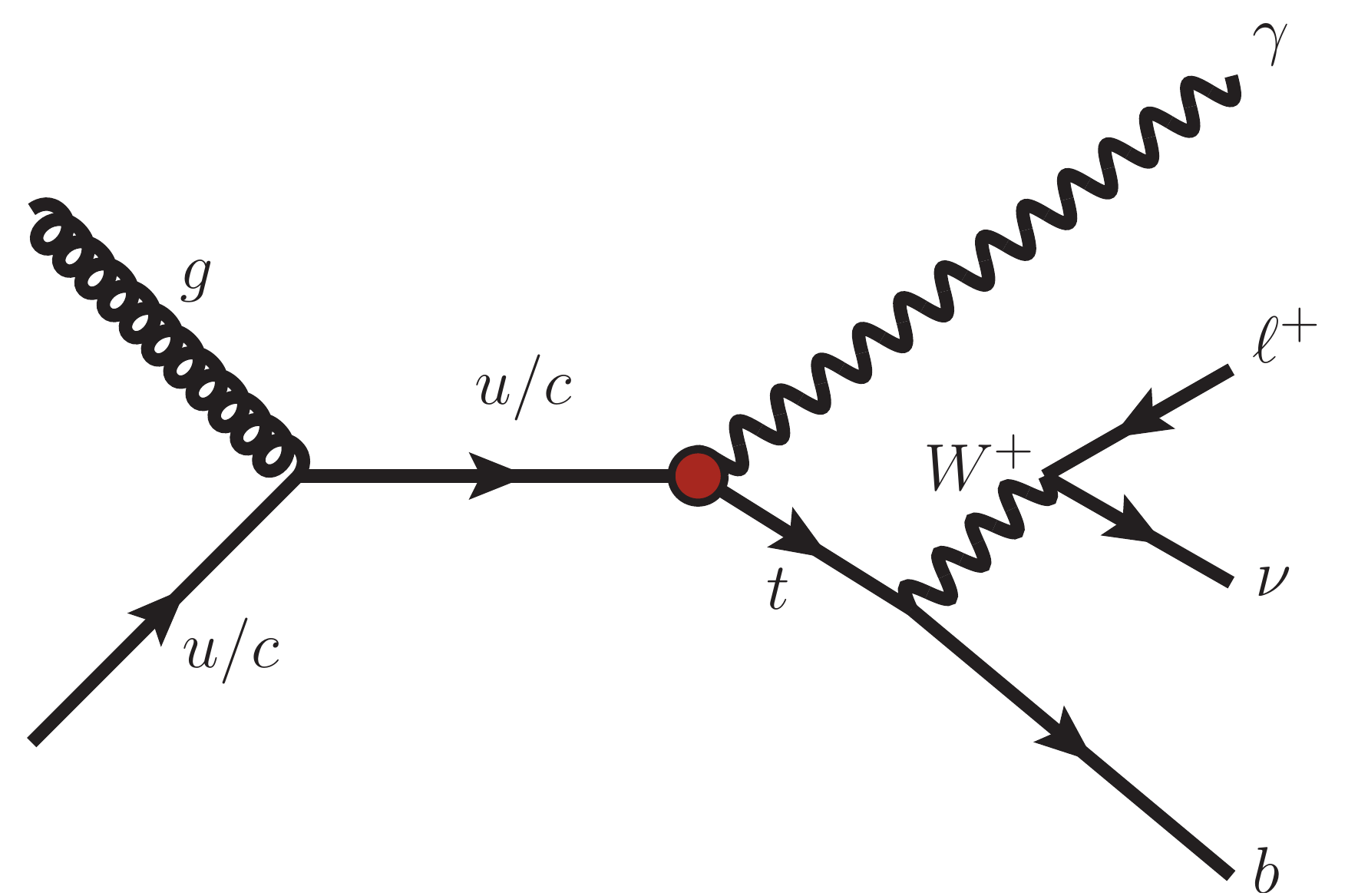} &
\includegraphics[width=0.49\columnwidth]{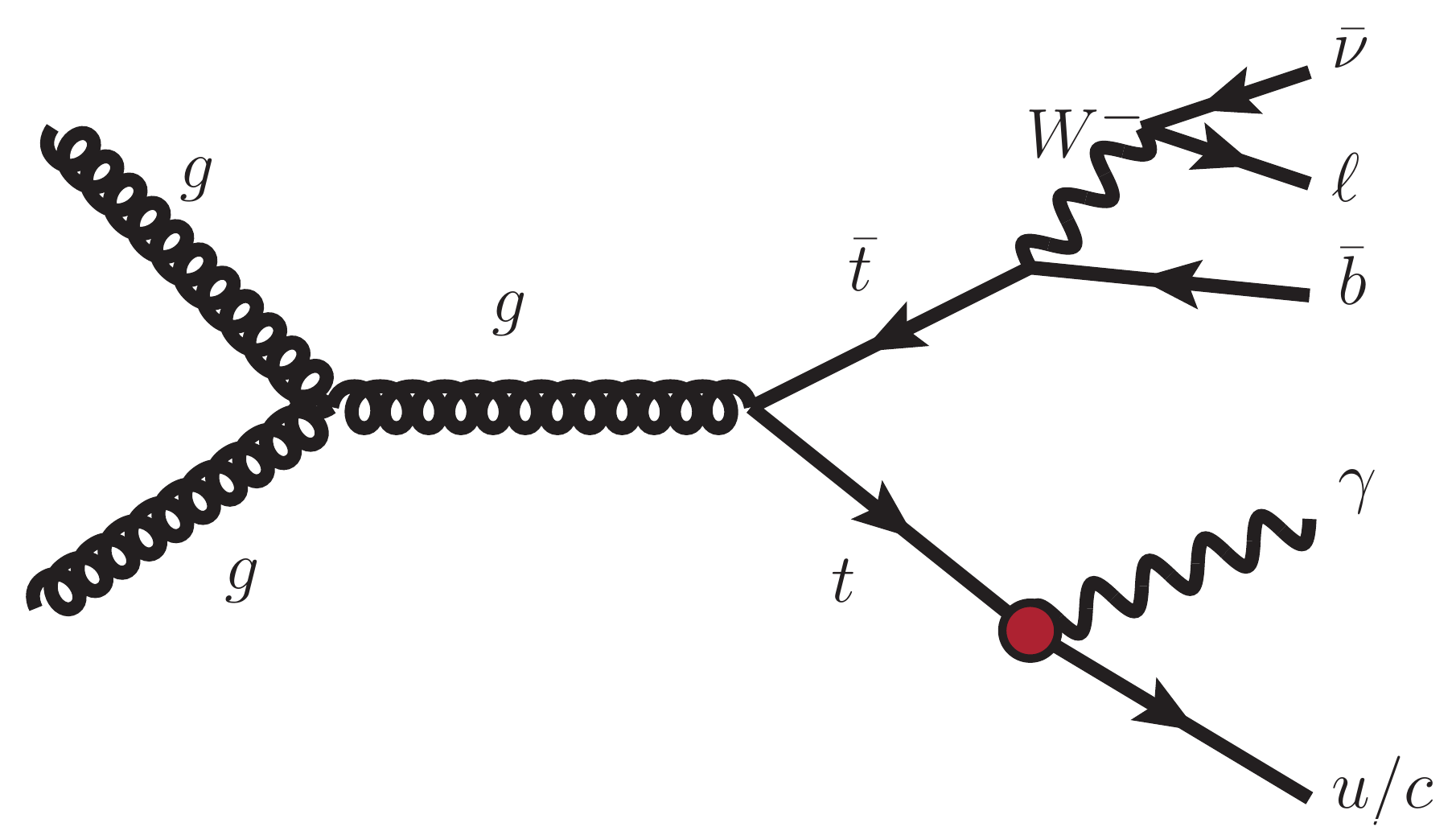}
\end{tabular}
\caption{Tree-level Feynman diagrams for top-quark production (left) and decay (right) via FCNCs. The $tq\gamma$ vertex, which is not present in the SM, is highlighted.
\label{fig:feynman}}
\end{center}
\end{figure}

This Letter presents a search for FCNCs in processes with a top quark ($t$) and a photon ($\gamma$) based on data collected with the ATLAS experiment at $\sqrt{s}$ = 13~\TeV. This analysis is most sensitive to the production of a single top quark plus a photon, but also considers the decay of a pair-produced top quark into an up or charm quark ($q$) plus a photon. Tree-level Feynman diagrams for these processes are shown in Figure~\ref{fig:feynman}, where in both cases, exactly one top quark decays via the SM-favoured $tWb$ coupling. Compared to the SM production of a top quark and a photon, the FCNC processes result in higher photon transverse momenta on average.
 
FCNC contributions to the decay mode ($t\rightarrow q \gamma$) and the production mode ($q \rightarrow t \gamma$) can be expressed in terms of effective coupling parameters but also in terms of branching ratios and cross sections~\cite{AguilarSaavedra:2008zc,Durieux:2014xla}.
In the former case and following the notation in Ref.~\cite{Grzadkowski:2010es}, the corresponding operators are $O_{\mathrm{uB}}^{(ij)}$ and $O_{\mathrm{uW}}^{(ij)}$, where $i\neq j$ are indices for the quark generation. In general, left-handed (LH) and right-handed (RH) couplings could exist, which result in different kinematic properties of the top-quark decay products, such as the transverse momentum of the charged lepton in semileptonic top-quark decays. The most stringent limits to date are limits on branching ratios of $\mathcal{B}(t \rightarrow u \, \gamma) < 1.3 \cdot 10^{-4}$ and $\mathcal{B}(t \rightarrow c \, \gamma) < 1.7 \cdot 10^{-3}$ set by the CMS Collaboration, assuming equal left- and right-handed couplings~\cite{Khachatryan:2015att}.
 
\section{ATLAS detector}
The ATLAS experiment~\cite{PERF-2007-01} at the LHC is a multi-purpose particle detector with a forward-backward symmetric cylindrical geometry and a near \(4\pi\) coverage in solid angle.\footnote{\AtlasCoordFootnote} It consists of an inner tracking detector~(ID) surrounded by a thin superconducting solenoid providing a \SI{2}{\tesla} axial magnetic field, electromagnetic and hadron calorimeters, and a muon spectrometer~(MS). The ID covers the pseudorapidity range \(|\eta| < 2.5\). It consists of silicon pixel, silicon microstrip, and transition radiation tracking detectors. Lead/liquid-argon (LAr) sampling calorimeters provide electromagnetic (EM) energy measurements with high granularity. A hadron (steel/scintillator-tile) calorimeter covers the central pseudorapidity range (\(|\eta| < 1.7\)). The end-cap and forward regions are instrumented with LAr calorimeters for both EM and hadronic energy measurements up to \(|\eta| = 4.9\). The MS surrounds the calorimeters and is based on three large air-core toroidal superconducting magnets with eight coils each. The field integral of the toroids ranges between \num{2.0} and \SI{6.0}{\tesla\metre} across most of the detector. The MS includes a system of precision tracking chambers and fast detectors for triggering. A two-level trigger system is used to select events. The first-level trigger is implemented in hardware and uses a subset of the detector information to reduce the accepted rate to at most nearly \SI{100}{\kHz}. This is followed by a software-based trigger that reduces the accepted event rate to \SI{1}{\kHz} on average depending on the data-taking conditions.
 
\section{Analysis strategy}
The search strategy selects events with a final state containing one prompt photon and the decay products of a top quark, namely an electron or a muon, a $b$-tagged jet, and missing transverse momentum, and estimates contributions from FCNC processes in this background-dominated dataset. A signal region (SR) is defined by loose requirements on the kinematic properties of the final-state objects, giving rise to a large acceptance for signal events in the production mode. With this selection, the search is most sensitive to FCNCs in this mode, but the decay mode is included in the analysis. The main background contribution stems from electrons misidentified as photons, primarily in top-quark--anti-top-quark events (\ttbar). These contributions and contributions from hadrons misidentified as photons (both labelled ``fakes'' in the following) are modelled by Monte Carlo (MC) simulations and scaled to data-driven estimates.  Photons produced in association with a leptonically decaying $W$ or $Z$ boson are estimated in control regions (CRs) which do not overlap with the SR but are kinematically similar to it. The predictions for other small prompt-photon background processes are taken from MC simulation and include $\ttbar\gamma$ production, single-top quark production in association with a photon and the production of two massive gauge bosons with a photon. As the latter two processes result in a small contribution to the total background prediction, it is sufficient that in these processes prompt photons are generated by the parton-shower program. Signal and background events are distinguished using a neural network (NN). Finally, the signal contribution is estimated in a profile likelihood fit to the NN output distributions in the SR and the CRs, in which each source of systematic uncertainty is modelled as a nuisance parameter.
 
\section{Data and simulation}
\label{sec:samples}
The proton--proton ($pp$) collision data analysed were recorded with the ATLAS detector from 2015 to 2017 at a centre-of-mass energy of $\sqrt{s} = 13$~\TeV. The average number of interactions per bunch crossing was 13.4, 25.1, and 37.8 in 2015, 2016 and 2017, respectively. Events were selected by single-lepton triggers~\cite{TRIG-2016-01} and required to have at least one reconstructed primary vertex with at least three assigned tracks that have a transverse momentum greater than 400~\MeV. After the application of data-quality requirements, the data sample corresponds to an integrated luminosity of 81\,fb$^{-1}$. It is obtained using the LUCID-2 detector~\cite{LUCID2} for the primary luminosity measurements.
 
The data were modelled by MC simulations of the signal and background processes. After event generation, the response of the ATLAS detector was simulated using \GEANT4~\cite{Agostinelli:2002hh} with a full detector model~\cite{SOFT-2010-01} or modelled by a fast simulation~\cite{ATL-PHYS-PUB-2010-013}. To account for additional $pp$ collisions (pile-up), inelastic $pp$ interactions were superimposed on the hard-scattering events and weighted according to the observed pile-up distribution. The pile-up events were simulated using \PYTHIAV{8.186}~\cite{Sjostrand:2014zea}, with the A3~\cite{ATL-PHYS-PUB-2016-017} set of tuned parameters (A3 tune).
 
The simulated signal samples were generated using \MGMCatNLOV{2.4.3}~\cite{Alwall:2014hca} with the TopFCNC model~\cite{Degrande:2014tta,Durieux:2014xla} at next-to-leading order (NLO) in QCD and the NNPDF3.0NLO~\cite{Ball:2014uwa} set of parton distribution functions (PDFs). The parton showering was done with \PYTHIAV{8.212} with the A14 tune set~\cite{ATL-PHYS-PUB-2014-021}. Simulated samples of SM \ttbar and single-top-quark events were generated using \POWHEGBOX~\cite{Frixione:2007nw,Nason:2004rx,Frixione:2007vw,Alioli:2010xd,Re:2010bp,Frederix:2012dh,Alioli:2009je} with the NNPDF3.0NLO PDF set. The parton showering, hadronisation, and the underlying event were modelled using \PYTHIAV{8.230}. The top-quark mass $m_{\mathrm{top}}$ was set to 172.5~\GeV\ in these samples, and the $h_\mathrm{damp}$ parameter that controls the transverse momentum of the first gluon emitted was set to 1.5 times $m_{\mathrm{top}}$. Samples of $\ttbar \gamma$ events were generated as $2\rightarrow 7$ process at leading order using \MGMCatNLOV{2.3.3} and the NNPDF2.3LO PDF~\cite{Ball:2012cx} set and with the following fiducial photon criteria~\cite{TOPQ-2017-14}: photon $\pt > 15~\GeV$ and $|\eta|<5.0$, charged-lepton $\pt > 15~\GeV$ and $|\eta|<5.0$, and $\Delta R < 0.2$ between the photon and any charged final-state particle. The cross sections for SM \ttbar\ and single-top-quark production are scaled to the NNLO+NNLL predictions~\cite{Czakon:2011xx,Kidonakis:2011wy,Kidonakis:2010tc,Kidonakis:2010ux}. The leading-order cross section for $\ttbar \gamma$ production of 4.62~pb~\cite{TOPQ-2017-14} is scaled to the NLO predictions~\cite{Melnikov:2011ta} using a $k$-factor of 1.24. The NNLO+NNLL cross section for SM \ttbar\ production is also used to normalise the signal in the decay mode, using the corresponding FCNC branching ratio. The cross sections for the signal in the production mode, however, are calculated at NLO with \MGMCatNLO.
 
For the study of systematic uncertainties in the modelling of processes involving top quarks, simulated \ttbar\ samples were produced with \POWHEGBOX+~\HERWIGV{7.0.4}~\cite{Bahr:2008pv,Bellm:2015jjp} and \MGMCatNLOV{2.6.0} plus \PYTHIAV{8.212}. The MMHT2014LO~\cite{Harland-Lang:2014zoa} PDF set is used together with the H7-UE-MMHT~\cite{Bellm:2015jjp} tune. An additional \ttbar sample was produced with $h_\mathrm{damp}$ set to three times $m_{\mathrm{top}}$ and the factorisation and renormalisation scales set to half their nominal values using the A14 tune. For the $tW$ process, a  sample was produced with an alternative scheme for removing the overlap with $\ttbar$ production~\cite{Frixione:2008yi}. Moreover, single-top-quark samples were produced with \POWHEGBOX+~\HERWIGV{7.0.4} and \MGMCatNLOV{2.6.2} plus \PYTHIAV{8.212}. The NNPDF2.3LO PDF set is used as well as the A14 tune.
 
Processes with one or two heavy gauge bosons, in particular the processes $W$+$\gamma$+jets and $Z$+$\gamma$+jets, were simulated using \SHERPA~2.2.1 and 2.2.2~\cite{Gleisberg:2008ta} with the matrix elements calculated using Comix~\cite{Gleisberg:2008fv} and OpenLoops~\cite{Cascioli:2011va}. All matrix elements were merged with the \SHERPA parton showering~\cite{Schumann:2007mg} according to the ME+PS@NLO~\cite{Hoeche:2009rj} prescription. The NNPDF3.0NNLO PDF set was used.
 
An overlap removal scheme was applied to remove double-counting of events stemming from photon radiation in samples in which a photon was not explicitly required in the final state~\cite{TOPQ-2017-14}. This applies to the processes $\ttbar$, $W$+jets and $Z$+jets in order to remove the overlap with the $\ttbar\gamma$, $W$+$\gamma$+jets and $Z$+$\gamma$+jets samples.
 
Mismodelling of the photon $\pt$ distribution is observed in the $W$+$\gamma$+jet and $Z$+$\gamma$ CRs, which are defined in Section~\ref{sec:selection}. The photon $\pt$ spectrum in the $W$+$\gamma$+jets and $Z$+$\gamma$+jets processes was corrected by adjusting the MC prediction to the data in five $\pt$ bins using a linear function that only changes the shape of the distribution and not its normalisation. This correction to the photon $\pt$ improves the modelling of the NN output distribution in the CRs.
 
As discussed in Sections~\ref{sec:electronfakes} and~\ref{sec:hadronfakes}, the contribution of events with electrons and hadrons misidentified as photons is corrected using data. The contribution from processes with hadrons misidentified as leptons is estimated to be negligible.
 
\section{Object and event selection}
\label{sec:selection}
Electrons are reconstructed from clusters of energy deposits in the electromagnetic calorimeter cells with a matched ID track~\cite{EGAM-2018-01}. They are required to meet the \textit{tight} identification criteria~\cite{EGAM-2018-01}, and their tracks must point to the primary vertex. They must have a transverse momentum $\pt$ larger than 27~\GeV\ and $|\eta_{\mathrm{cluster}}|<2.47$, excluding $1.37 < |\eta_{\mathrm{cluster}}| < 1.52$, where $|\eta_{\mathrm{cluster}}|$ is the pseudorapidity of the electron's energy cluster. Muons are reconstructed by combining a track in the MS with a track in the ID~\cite{PERF-2015-10}. They are required to meet the \textit{medium} identification criteria~\cite{PERF-2015-10} and must point to the primary vertex. They must have $\pt > 27~\GeV$ and $|\eta|<2.5$. Isolated electrons and muons are selected by requiring the amount of energy in nearby energy deposits in the calorimeters and the scalar sum of the transverse momenta of nearby tracks in the ID to be small.
 
Photons are reconstructed from clusters of energy deposits in the electromagnetic calorimeter cells with no matched ID track (unconverted photons) or with one or two matched ID tracks that are compatible with the tracks from an electron or positron from a photon conversion (converted photons)~\cite{EGAM-2018-01}. They must have $\pt > 20~\GeV$ and $|\eta_{\mathrm{cluster}}|<2.37$, excluding $1.37 < |\eta_{\mathrm{cluster}}| < 1.52$. They are required to meet the \textit{tight} identification criteria for the shape of the shower in the electromagnetic calorimeter (shower shape) and for the energy deposited in the hadronic calorimeter~\cite{EGAM-2018-01}. Photons must be isolated from nearby energy deposits in the calorimeter and nearby tracks in the ID: the sum of the energy deposited ($\pt$ of the tracks) within $\Delta R = 0.4$ ($\Delta R = 0.2$) of the photon direction is required to be smaller than $0.022\times\pt + 2.45~\GeV$\linebreak ($0.065\times\pt$), excluding the photon energy deposition (tracks associated with the photon).
 
Jets are reconstructed from topological clusters~\cite{ATL-PHYS-PUB-2015-036,PERF-2014-07} in the calorimeters with the anti-$k_t$ algorithm~\cite{Cacciari:2008gp} using FastJet~\cite{Cacciari:2011ma} and a radius parameter of 0.4. Their energy is calibrated~\cite{PERF-2016-04}, and they must fulfil $\pt > 25~\GeV$ and $|\eta|<2.5$. Jets with $\pt < 120~\GeV$ and $|\eta|<2.4$ are required to pass a requirement on the jet-vertex-tagger (JVT)~\cite{PERF-2014-03} to suppress pile-up jets. Jets are $b$-tagged with the MV2c10 algorithm~\cite{FTAG-2018-01}, which uses a boosted decision tree with several $b$-tagging algorithms as input. The $b$-tagging efficiency for jets that originate from the hadronisation of $b$-quarks is 60\% in simulated $\ttbar$ events. The $b$-tagging rejection\footnote{The rejection is defined as the inverse of the efficiency.} for jets that originate from the hadronisation of $c$-quarks ($u$-, $d$-, $s$-quarks or gluons) is 23 (1200).
 
The magnitude of the missing transverse momentum \met\ is reconstructed from the vector sum of the $\pt$ of leptons, photons, and jets, as well as ID tracks that point to the primary vertex but are not associated with a reconstructed object (soft term)~\cite{PERF-2016-07}.
 
To avoid double-counting, objects are removed in the following order:
electrons sharing a track with a muon;
jets within $\Delta R = 0.2$ of an electron;
electrons within $\Delta R = 0.4$ of a jet;
jets within $\Delta R = 0.4$ of a muon if they have at most two associated tracks;
muons within $\Delta R = 0.4$ of a jet;
photons within $\Delta R = 0.4$ of an electron or muon;
jets within $\Delta R = 0.4$ of a photon.
 
Scale factors (SFs) are used to correct the efficiencies in simulation in order to match the efficiencies measured in data for the electron~\cite{EGAM-2018-01} and muon~\cite{PERF-2015-10} trigger, reconstruction, identification, and isolation criteria, as well as for the photon identification~\cite{EGAM-2018-01} and isolation requirements. SFs are also applied for the JVT requirement and for the $b$-tagging efficiencies for jets that originate from the hadronisation of $b$-quarks~\cite{PERF-2016-05}, $c$-quarks~\cite{FTAG-2018-01}, and $u$-, $d$-, $s$-quarks or gluons~\cite{ATLAS-CONF-2018-006}.
 
The selected events have exactly one electron or muon, exactly one photon, exactly one $b$-tagged jet and no further jets, and $\met > 30~\GeV$. This selection defines the SR with signal efficiencies\footnote{Here, the signal efficiency includes the signal loss due to the limited acceptance of the detector.} for the production mode of 3.03\% (2.45\%) for the LH (RH) $tu\gamma$ coupling and of 3.79\% (3.14\%) for the LH (RH) $tc\gamma$ coupling. These efficiencies are defined with respect to signal events that include a leptonic decay of the $W$ boson, i.e.\ a decay with an electron, muon or tau lepton. The efficiencies for the couplings that involve a $c$-quark are larger than those that involve a $u$-quark because of the difference in the kinematics that arises from the difference between the $u$- and $c$-quark PDFs. The difference between the efficiencies for the LH and the RH couplings is due to the kinematic distributions of the $W$ boson's decay products, which differ depending on the handedness of the top quark. The efficiencies for the decay mode are 0.45\% and 0.51\% for the $tu\gamma$ and the $tc\gamma$ coupling, respectively, and are lower due to the requirement of not more than one jet in the final state, i.e.\ this analysis is optimised for the production mode. The absolute statistical uncertainties in the efficiencies are 0.03\% or smaller. In the SR, 9557 data events are selected. The ratio of production-mode to decay-mode event yields is 4.2 (5.3) for the LH (RH) $tu\gamma$ coupling and it is 0.86 (0.68) for the LH (RH) $tc\gamma$ coupling, i.e.\ in the case of $tu\gamma$ coupling, the dominant signal process is indeed the production mode. However, in the case of the $tc\gamma$ coupling, the decay mode also plays an important role, because the production mode is suppressed by the $c$-quark PDF.
 
Two CRs are defined for the $W$+$\gamma$+jets and $Z$+$\gamma$+jets processes, which are dominated by the respective background process and kinematically close to the SR. The $W$+$\gamma$+jet CR is defined by the same criteria as the SR with two modifications: the jet must not be $b$-tagged and the lepton--photon invariant mass must be outside the range 60--100~\GeV\ to suppress the contribution from $Z$+jets events with one electron that is misidentified as a photon. The $Z$+$\gamma$ CR is defined by requiring exactly one photon and exactly two leptons with the same flavour but opposite electric charge. No requirement is made in the $Z$+$\gamma$ CR on the number of jets or \met. In the $W$+$\gamma$+jet CR, a total number of 127\,864 events are observed, and in the $Z$+$\gamma$ CR, the total number is 85\,347.
 
\section{Data-driven estimate of electrons misidentified as photons}
\label{sec:electronfakes}
Electrons can be misidentified as photons, for example, if the track of the electron is not reconstructed or if the matching criteria between the track and cluster are not met. In particular, dileptonic $\ttbar$ events with at least one electron in the final state can enter the SR if an electron is misidentified as a photon. The probability for an electron to be misidentified as a photon, \egfake, is measured from data and simulation following the methodology used previously~\cite{PERF-2017-02}, and a SF is derived that is applied to the simulation.
 
Two regions are defined to measure \egfake, called ``electron fake regions'' (EFR) in the following. The $Z\rightarrow e\gamma$ ($Z\rightarrow ee$) EFR is defined by requiring exactly one electron and one photon (exactly two electrons with opposite electric charge and no photons) with an electron--photon (dielectron) invariant mass in the range 60--120~\GeV, a veto on the presence of jets, and $\met < 30~\GeV$. Neither EFR overlaps with the SR nor the CRs. The $Z\rightarrow e\gamma$ EFR is rich in $Z\rightarrow ee$ events with one electron misidentified as a photon, and the $Z\rightarrow ee$ EFR is rich in $Z\rightarrow ee$ events.
 
In the $Z\rightarrow ee$ and the $Z\rightarrow e\gamma$ EFRs, the dielectron invariant mass or the electron--photon invariant mass, respectively, is fitted with analytic signal (for $Z\rightarrow ee$ with both electrons correctly identified or with one electron misidentified as a photon, respectively) and background functions. The signal function is a double-sided Crystal Ball function, and the background function is a fourth-order Bernstein polynomial. The integrals of the aforementioned fitted signal functions are divided in order to estimate $2 \egfake$, where the factor of two accounts for the two electrons in $Z\rightarrow ee$ events that may be misidentified as a photon. In the $Z\rightarrow e\gamma$ EFR, the expected contribution from $Z\rightarrow ee\gamma$ events, relative to the signal, is 8.8\% and it is subtracted from the integral, because this process mainly contributes to events with prompt photons and in which one electron was not reconstructed or did not pass the identification or isolation criteria.
 
Systematic uncertainties from several sources are evaluated: the range of the invariant-mass fit is changed from 60--120~\GeV\ to 65--115~\GeV; the parameters of the signal function, except for its normalisation, are set to the values extracted from simulation and a Gaussian function is used for the background; instead of subtracting the expected relative contribution from $Z\rightarrow ee\gamma$ events in the $Z\rightarrow e\gamma$ EFR, the expected absolute contribution is subtracted. For each of these variations, a systematic uncertainty for \egfake\ is estimated as the deviation from the nominal value. The largest effect is due to the variation of the signal and background functions. The value of \egfake\ is $3.11\% \pm 0.01\%\,\left(\mathrm{stat.}\right)\pm 0.13\%\,\left(\mathrm{syst.}\right)$. The SF for the simulation is $0.978\pm 0.004\,\left(\mathrm{stat.}\right)\pm 0.040\,\left(\mathrm{syst.}\right)$.
 
The modelling of kinematic variables is checked in a validation region. The event selection for this validation region is similar to the SR selection, but a few changes are made in order to enhance the contributions from $Z\rightarrow ee$ events and dileptonic \ttbar\ events with a misidentified electron, while ensuring no overlap with the SR, the CRs, or the EFRs. The validation region is defined by requiring exactly one photon and one electron with an invariant mass in the range 70--110~\GeV, at least one jet, and $\met<30~\GeV$. Satisfactory modelling of the kinematic variables is observed, but the relative uncertainty in \egfake\ is increased to 10\% in order to cover the difference in the normalisation observed between data and the prediction.
 
\section{Data-driven estimate of hadrons misidentified as photons}
\label{sec:hadronfakes}
In some cases, hadrons can be misidentified as photons. For example, this can happen when a meson decays into two photons that are reconstructed as a single cluster in the electromagnetic calorimeter. Processes such as $\ttbar$ production can enter the SR if a high-energy hadron is misidentified as a photon. The number of events with misidentified hadrons is estimated from data, and a SF is applied to the simulation, defined as the estimated number of events in data divided by the predicted number in simulation. The SF is only used to correct the overall normalisation of this background, and the shapes of kinematic distributions are taken from simulations with the associated systematic uncertainties.
 
Three hadron fake regions (HFR) are defined by the same criteria as the SR but with modified photon criteria: $\mathrm{HFR}^{\mathrm{pass|fail}}$, $\mathrm{HFR}^{\mathrm{fail|pass}}$, and $\mathrm{HFR}^{\mathrm{fail|fail}}$. If the first index is ``pass'', the photon has met the identification criteria defined in Section~\ref{sec:selection}. If the first index is ``fail'', the photon has failed to meet at least one of the criteria on the shower shapes that are calculated from the finely-segmented first layer of the electromagnetic calorimeter; however, it is required to meet all tight photon-identification criteria for the other shower variables. The second index represents whether the photon meets or fails to meet the isolation criterion.
 
Only the first-layer shower shapes are considered for the first index because these are mostly sensitive to the core of the shower and are expected to be only weakly correlated with the isolation variables, which are sensitive to the energy surrounding the photon. The number of SR events with misidentified hadrons is estimated as $N(\mathrm{HFR}^{\mathrm{pass|fail}})\times N(\mathrm{HFR}^{\mathrm{fail|pass}}) / N(\mathrm{HFR}^{\mathrm{fail|fail}})$, where $N$ is the number of observed events after subtracting both the expected number of events with misidentified electrons and the expected fraction of events with prompt photons (leakage). This estimate is additionally corrected for the non-zero correlation between the criteria for the first-layer shower shapes and the isolation criterion. The correction factor is determined using MC simulations and amounts to 0.85 with a statistical uncertainty of 0.14.
 
Systematic uncertainties from several sources are evaluated: the correction factor for the non-zero correlation is conservatively varied by $\pm 50\%$; the SF for misidentified electrons, used for the subtraction discussed above, is varied by one standard deviation up and down ($\pm 1\sigma$), and the larger of the two deviations is considered as a systematic uncertainty; and  instead of subtracting the expected fraction of events with prompt photons, the expected prompt-photon contribution is subtracted in each HFR. For each of these variations, a systematic uncertainty is estimated as the resulting deviation from the nominal value. The largest effect is the variation of the correction factor. The SF for the simulations is $1.7\pm 0.3\,\left(\mathrm{stat.}\right)\pm 1.0\,\left(\mathrm{syst.}\right)$.
 
\section{Neural network for discrimination between signal and background}
The signal is distinguished from the sum of the background processes by a fully connected feed-forward neural network (NN) with backpropagation, implemented in {\textsc Keras}~\cite{chollet2015keras} with the {\textsc TensorFlow}~\cite{tensorflow2015-whitepaper} back end. Separate NNs are trained for FCNC processes with a $tu\gamma$ or a $tc\gamma$ vertex and with LH or RH couplings. For the signal sample, the production mode was chosen, since the kinematic differences between the production mode and the background are more pronounced than for the decay mode and thus lead to better discrimination between signal and background.
 
Ten variables are inputs to the NN: the \pt\ of the photon, the lepton and the jet; the charge of the lepton; \met; the lepton--photon and lepton--jet invariant masses; the $\Delta R$ between the lepton and the photon, between the lepton and the jet, and between the jet and the photon. Additional kinematic variables were tested and did not improve the discrimination between signal and background. All variables are transformed using {\textsc scikit-learn}'s~\cite{Pedregosa:2012toh} RobustScaler.
 
The NN is trained with the {\textsc Adam} optimiser~\cite{DBLP:journals/corr/KingmaB14}. The MC samples are divided into 80\% for training and 20\% for testing, so that approximately 63\,000 background MC events and approximately 10\,000-13\,000 signal MC events are available for the training---depending on the coupling. Two hidden layers with 11 nodes each are used and the hyperparameters of the NN are chosen from a series of tested values in a procedure with threefold cross validation.
 
\section{Systematic uncertainties}
Systematic effects may change the expected numbers of events from the signal and background processes and the shape of the fitted discriminants in the SR and in the CRs. These effects are evaluated by varying each source of systematic uncertainty by $\pm1\sigma$ and considering the resulting difference from the nominal expectation as the uncertainty. For some sources, only one variation is available and the difference is symmetrised using the full difference. For sources with two variations, their effects are symmetrised using the average difference from the nominal prediction.
 
Uncertainties due to the theoretical cross sections are evaluated by varying them by $\pm 5.6\%$ for $\ttbar$ production~\cite{Czakon:2011xx,Botje:2011sn,Martin:2009bu,Gao:2013xoa,Ball:2012cx}, by $\pm 8\%$ for $\ttbar$+$\gamma$ production~\cite{Melnikov:2011ta}, by $^{+4.0}_{-3.4}\%$ ($^{+5.0}_{-4.5}\%$) for $t$-channel single-$t$ (single-$\bar{t}$) production~\cite{Kidonakis:2011wy}, by $^{+3.6}_{-3.1}\%$ ($^{+4.8}_{-4.3}\%$) for $s$-channel single-$t$ (single-$\bar{t}$) production~\cite{Kidonakis:2010tc}, by $\pm 5.3\%$ for $tW$ production~\cite{Kidonakis:2010ux}, by $\pm 5\%$ for $W$+jets and $Z$+jets production~\cite{ATL-PHYS-PUB-2016-003}, and by $\pm 6\%$ for diboson production~\cite{ATL-PHYS-PUB-2016-002}. No cross-section uncertainty is considered for $W$+$\gamma$+jets and $Z$+$\gamma$+jets production, because their normalisations are determined in the fit.
 
Uncertainties due to the modelling of the signal are estimated by considering variations of the renormalisation and factorisation scales by factors of 2 and 0.5, but normalising the signal to the nominal cross section. In each bin of a distribution, the largest deviation among all variations is considered as an uncertainty (envelope). In addition, uncertainties due to the PDFs are estimated by following the PDF4LHC prescription for Run~2~\cite{Butterworth:2015oua}.
 
Uncertainties due to the renormalisation and factorisation scales and from the PDFs of the background processes are estimated separately for each process following the same procedure as for the signal. For the \ttbar\ and single-top processes, however, the scale variations are already included in the estimation of the uncertainty in the modelling of the initial-state radiation (see below). For the $W$+$\gamma$+jets and the $Z$+$\gamma$+jets processes, a correction is applied to the photon $\pt$ spectrum, as described in Section~\ref{sec:samples}. To account for the uncertainty due to the photon $\pt$ correction, a conservative uncertainty is applied, for which the prediction with the correction applied is compared with the prediction without the correction.
 
For all background processes---except for $W$+$\gamma$+jets and $Z$+$\gamma$+jets production, for which the normalisation is estimated by a free parameter in the fit---an uncertainty of 2\% in the integrated luminosity is included~\cite{ATLAS-CONF-2019-021}. The uncertainty due to pile-up is determined by varying the average number of interactions by 9\% in the simulation. The uncertainties due to the SFs for electrons and hadrons that are misidentified as photons are determined as described in Sections~\ref{sec:electronfakes} and~\ref{sec:hadronfakes}.
 
To estimate the uncertainty due to the production of $W$ and $Z$ bosons together with $b$-quarks, the shape of the SR distribution in events with jets that originate from the hadronisation of a $b$-quark is used for events with jets that originate from other quarks or gluons (and vice versa) for the $W$+jets, $W$+$\gamma$+jets, $Z$+jets, $Z$+$\gamma$+jets, and diboson processes. Differences between the shapes of these backgrounds in association with $b$-quarks or with other quarks or gluons are small, however. An additional uncertainty in the normalisation of $W$+$\gamma$ production in association with $b$-quarks of 50\% is assigned, covering observed differences between data and predictions in measurements of $W$ and $Z$ bosons in association with $b$-quarks~\cite{CMS-SMP-14-020,CMS-SMP-14-010,STDM-2012-11,STDM-2012-15}.
 
To estimate the uncertainty due to the modelling of initial- and final-state radiation in $\ttbar$ and single-top-quark production, the effects of varying the A14 tune's parameter values is evaluated. In addition, for $\ttbar$ production, the sample generated with $h_{\mathrm{damp}} = 3m_{\mathrm{top}}$, the factorisation and renormalisation scales set to half their nominal values, and a variation of the A14 tune's parameter values is used to estimate the uncertainty due to the modelling of initial-state radiation. To estimate the uncertainty due to our choice of generator and shower programs for $\ttbar$ and single-top-quark production, the nominal MC samples, generated with \POWHEGBOX+~\PYTHIAV{8}, are replaced with samples generated with \MGMCatNLOV+~\PYTHIAV{8} and with \POWHEGBOX+~\HERWIGV{7}. To evaluate the uncertainty due to the scheme for removing the overlap with $\ttbar$ production for the $tW$ process, the nominal sample is compared with a sample produced with an alternative scheme~\cite{Frixione:2008yi}.
 
For the triggering, reconstruction, identification, and calibration of the objects, the following systematic uncertainties are evaluated: electron and muon triggers, reconstruction, identification and isolation SFs~\cite{EGAM-2018-01,PERF-2015-10}; photon identification~\cite{EGAM-2018-01} and isolation SFs; electron- and photon-energy and muon-momentum calibration and resolution~\cite{EGAM-2018-01,PERF-2015-10}; jet energy scale (JES)~\cite{PERF-2016-04} and jet energy resolution (JER)~\cite{ATL-PHYS-PUB-2015-015}; JVT SF; $b$-tagging SFs~\cite{FTAG-2018-01,ATLAS-CONF-2018-001,ATLAS-CONF-2018-006}; \met\ soft term~\cite{PERF-2016-07}.
 
\section{Results}
\begin{figure}[ht!]
\begin{center}
\includegraphics[width=0.49\columnwidth]{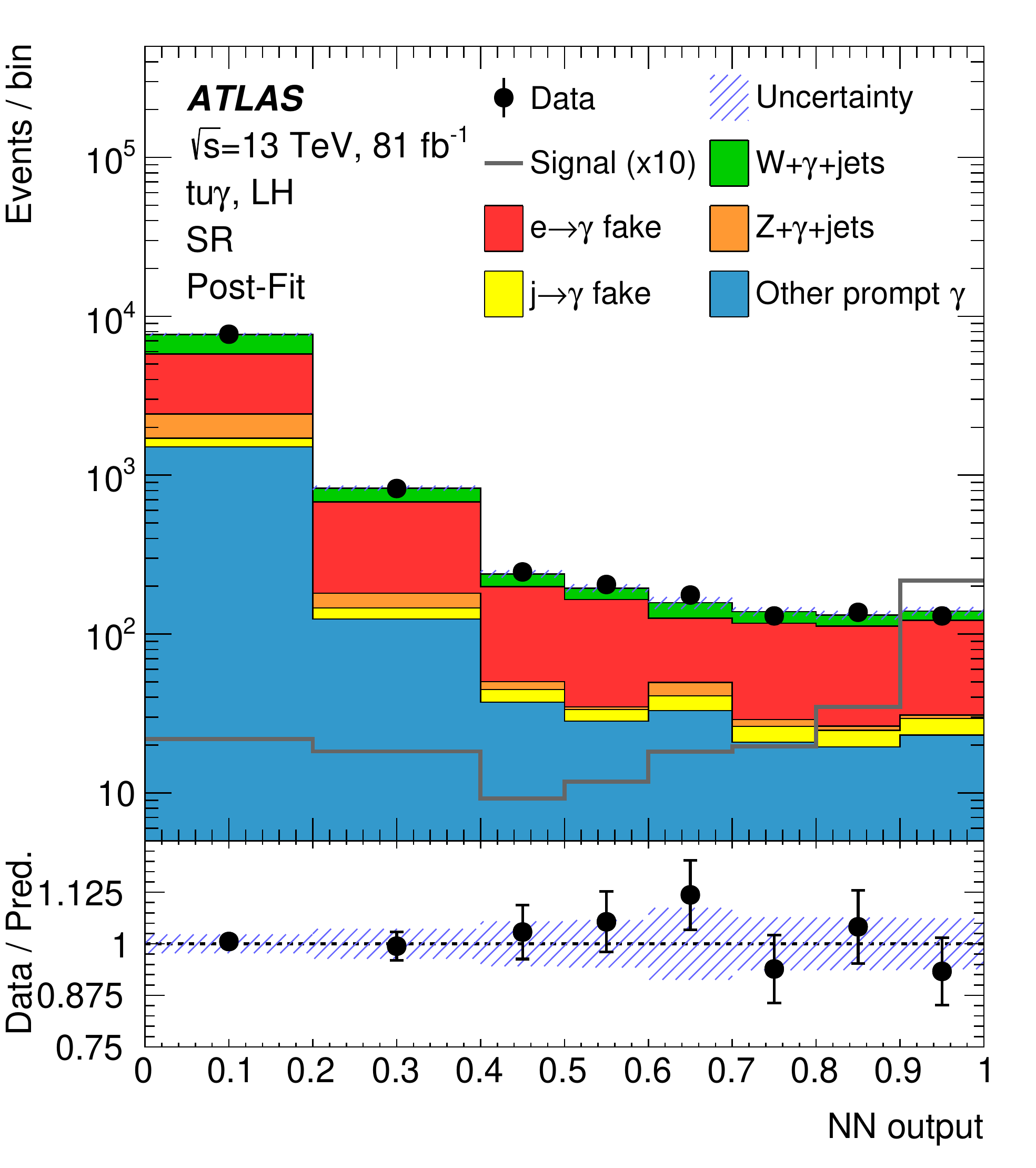} \\
\includegraphics[width=0.49\columnwidth]{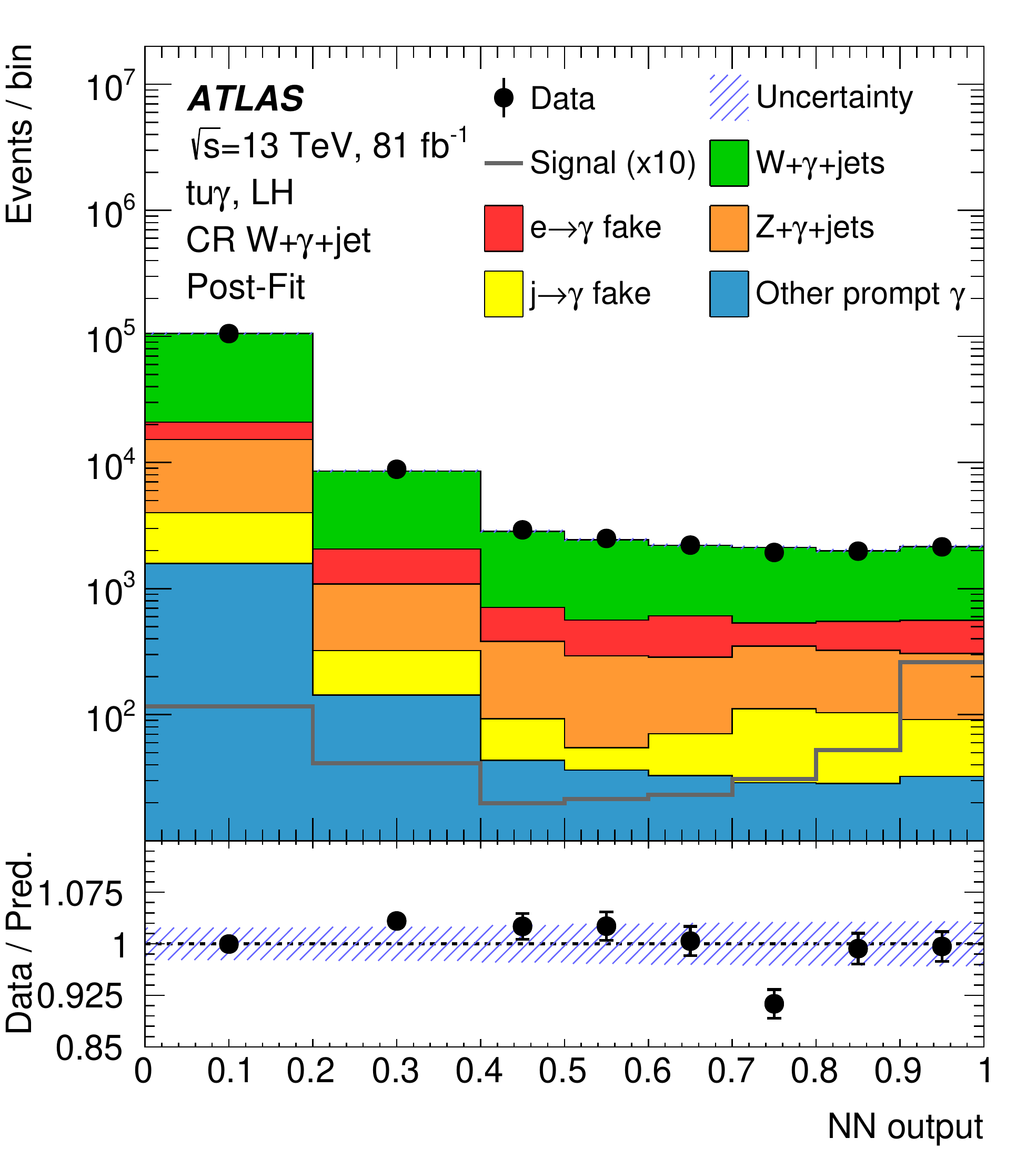}
\includegraphics[width=0.49\columnwidth]{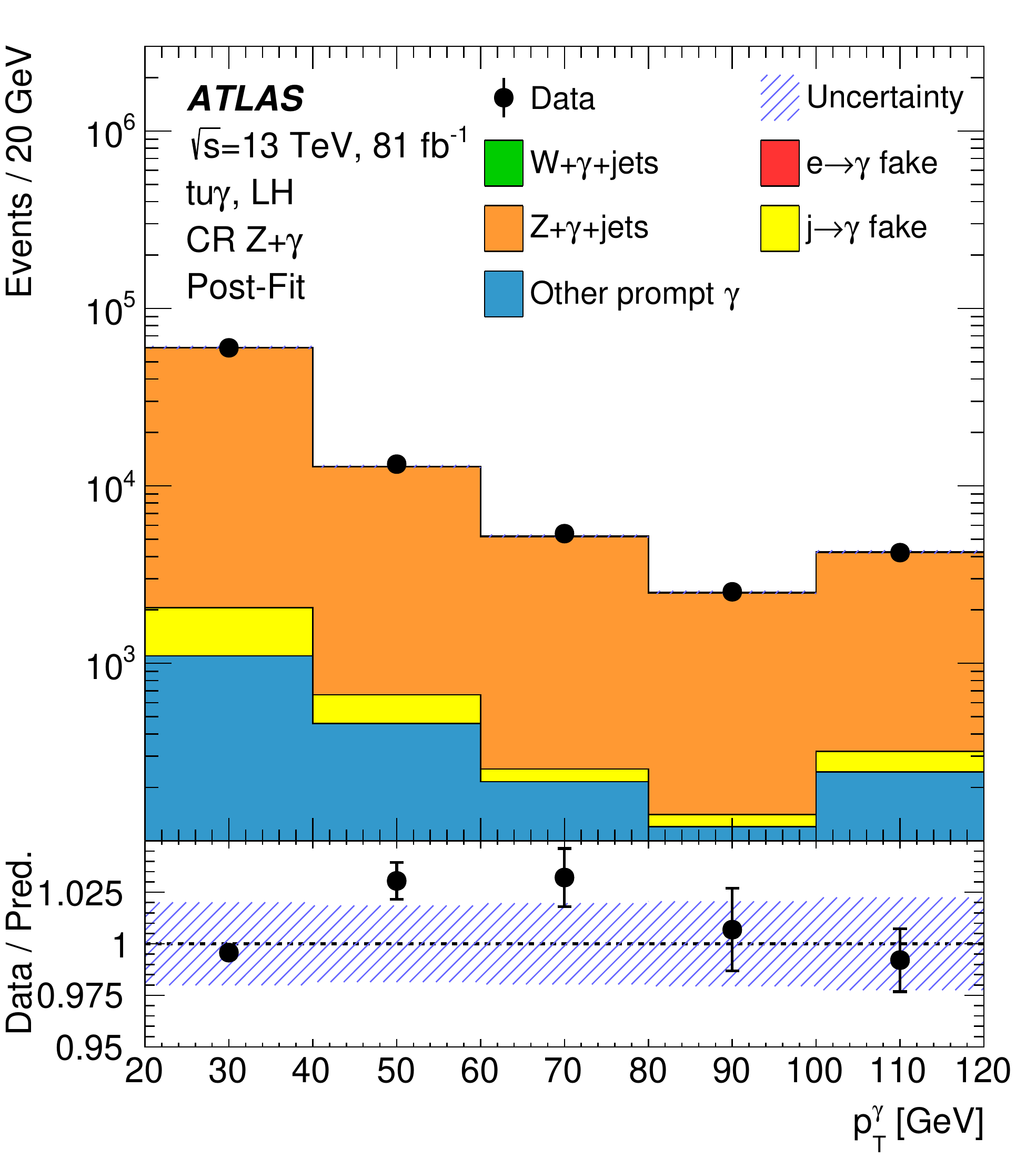}
\caption{Post-fit distributions of a background-only fit to the SR and the CRs of the NN output in the SR (top) and the $W$+$\gamma$+jet CR (bottom left) and of the \pt-distribution of the $Z$+$\gamma$ CR (bottom right). The last bin of the distribution in the $Z$+$\gamma$ CR contains the overflow. In addition, in the SR and in the $W$+$\gamma$+jet CR, the expected signal is overlaid for an effective coupling strength corresponding to the observed limit multiplied by a factor of ten. In the $Z$+$\gamma$ CR, the expected signal is not shown, because it is negligible.
\label{fig:postfit}}
\end{center}
\end{figure}

The normalisations of the signal contribution and the two contributions from $W$/$Z$+$\gamma$+jets production are obtained from a simultaneous binned profile-likelihood fit to the NN-output distribution of the SR and $W$+$\gamma$+jet CR as well as the photon-\pt\ distribution of the $Z$+$\gamma$ CR. The signal contribution scales the production- and decay-mode contributions consistently. Each source of systematic uncertainty is associated with a nuisance parameter. In Table~\ref{tab:yields}, the expected number of events after a background-only fit to the SR and CRs for the case of the LH $tu\gamma$ coupling are shown, as well as the observed number of events. Figure~\ref{fig:postfit} shows the corresponding post-fit distributions. The qualitative features of these distributions are similar for the other couplings studied.
 
\begin{table}[ht!]
\begin{center}
\caption{Expected number of events for the different background contributions in the SR and the two CRs after the fit including all uncertainties, as well as the observed number of events.}
\label{tab:yields}
\begin{tabular}{lR{1.5cm}@{\hspace{1.5pt}}c@{\hspace{1.5pt}}p{1cm}R{1.5cm}@{\hspace{1.5pt}}c@{\hspace{1.5pt}}p{1cm}R{1.5cm}@{\hspace{1.5pt}}c@{\hspace{1.5pt}}p{1cm}}
\hline
& \multicolumn{3}{c}{SR} & \multicolumn{3}{c}{$W$+$\gamma$+jet CR} & \multicolumn{3}{c}{$Z$+$\gamma$ CR}\\
\hline
$e\rightarrow\gamma$ fake & 4\,500 & $\pm$ & 400 & 8\,200 & $\pm$ & 1\,300 & 236 & $\pm$ & 32\\
$j\rightarrow\gamma$ fake & 260 & $\pm$ & 200 & 2\,900 & $\pm$ & 2\,000 & 1\,300 & $\pm$ & 1\,000\\
$Z$+$\gamma$+jets & 780 & $\pm$ & 100 & 13\,400 & $\pm$ & 1\,300 & 81\,400 & $\pm$ & 1\,900\\
$W$+$\gamma$+jets & 2\,200 & $\pm$ & 400 & 101\,200 & $\pm$ & 2\,800 & 6 & $\pm$ & 2\\
Other prompt $\gamma$ & 1\,800 & $\pm$ & 400 & 1\,900 & $\pm$ & 500 & 2\,140 & $\pm$ & 200\\
\hline
Total background & 9\,500 & $\pm$ & 220 & 127\,700 & $\pm$ & 3\,000  & 85\,100 & $\pm$ & 1\,600\\
Data & 9\,557 & & & 127\,864 & & & 85\,347 & & \\
\hline
\end{tabular}
\end{center}
\end{table}
 
The data and SM predictions agree within uncertainties and no significant FCNC contributions are observed. From the 95\% confidence level (CL) limits on the signal contribution, derived using the $CL_{s}$ method~\cite{Read:2002hq}, the corresponding limits on the effective coupling parameters are calculated, and from these the limits on the production cross section and branching ratios are calculated. The background contributions from photons produced in association with a leptonically decaying $W$ or $Z$ boson are scaled by normalisation factors estimated to be $1.25\pm0.09$ and $1.12\pm0.12$, respectively, from the fit for the LH $tu\gamma$ coupling. The normalisation values determined in the fits for the other couplings are similar. The observed and expected 95\% CL limits on the effective coupling strengths, the production cross section and the branching ratio are summarised in Table~\ref{tab:limits} for different vertices and couplings.  The sources of systematic uncertainty with the largest impact on the estimated signal contribution depend on the coupling studied. Among them are the jet energy resolution, the reweighting of the photon \pt, the factorisation and renormalisation scales, the choice of generator for the simulation of the \ttbar\ and single-top processes, and the uncertainties due to the limited number of Monte Carlo events. The resulting limits on the strength of the effective operators are complementary to current limits on the single operators from a search for an FCNC $tqZ$ coupling~\cite{TOPQ-2017-06}.
 
\begin{table}[ht!]
\begin{center}
\caption{Observed (expected) 95\% CL limits on the effective coupling strengths for different vertices and couplings, the production cross section, and the branching ratio. For the former, the energy scale is assumed to be $\Lambda$ = 1~\TeV.
\label{tab:limits}}
\begin{tabular}{ccccc}
\hline
Observable & Vertex & Coupling & Obs. & Exp. \\
\hline
$\left| C_{\mathrm{uW}}^{(13)*} + C_{\mathrm{uB}}^{(13)*} \right|$ & $tu\gamma$ & LH & 0.19 & $0.22^{\mathstrut{+0.04}}_{\mathstrut{-0.03}}$ \\
$\left| C_{\mathrm{uW}}^{(31)}  + C_{\mathrm{uB}}^{(31)} \right|$  & $tu\gamma$ & RH & 0.27 & $0.27^{\mathstrut{+0.05}}_{\mathstrut{-0.04}}$ \\
$\left| C_{\mathrm{uW}}^{(23)*} + C_{\mathrm{uB}}^{(23)*} \right|$ & $tc\gamma$ & LH & 0.52 & $0.57^{\mathstrut{+0.11}}_{\mathstrut{-0.09}}$ \\
$\left| C_{\mathrm{uW}}^{(32)}  + C_{\mathrm{uB}}^{(32)} \right|$  & $tc\gamma$ & RH & 0.48 & $0.59^{\mathstrut{+0.12}}_{\mathstrut{-0.09}}$ \\
\hline
$\sigma(pp\rightarrow t \gamma)$\, [fb] & $tu\gamma$ & LH & 36 & $52^{\mathstrut{+21}}_{\mathstrut{-14}}$ \\
$\sigma(pp\rightarrow t \gamma)$\, [fb] & $tu\gamma$ & RH & 78 & $75^{\mathstrut{+31}}_{\mathstrut{-21}}$ \\
$\sigma(pp\rightarrow t \gamma)$\, [fb] & $tc\gamma$ & LH & 40 & $49^{\mathstrut{+20}}_{\mathstrut{-14}}$ \\
$\sigma(pp\rightarrow t \gamma)$\, [fb] & $tc\gamma$ & RH & 33 & $52^{\mathstrut{+22}}_{\mathstrut{-14}}$ \\
\hline
$\mathcal{B}(t \rightarrow q \gamma)\, [10^{-5}]$ & $tu\gamma$ & LH & 2.8 & $4.0^{\mathstrut{+1.6}}_{\mathstrut{-1.1}}$ \\
$\mathcal{B}(t \rightarrow q \gamma)\, [10^{-5}]$ & $tu\gamma$ & RH & 6.1  & $5.9^{\mathstrut{+2.4}}_{\mathstrut{-1.6}}$ \\
$\mathcal{B}(t \rightarrow q \gamma)\, [10^{-5}]$ & $tc\gamma$ & LH & 22 & $27^{\mathstrut{+11}}_{\mathstrut{-7}}$ \\
$\mathcal{B}(t \rightarrow q \gamma)\, [10^{-5}]$ & $tc\gamma$ & RH & 18 & $28^{\mathstrut{+12}}_{\mathstrut{-8}}$ \\
\hline
\end{tabular}
\end{center}
\end{table}
 
\FloatBarrier
 
\section{Conclusion}
A search for flavour-changing neutral currents (FCNCs) in events with one top quark and a photon is presented using 81~fb$^{-1}$ of $\sqrt{s} = 13$~\TeV\ $pp$ data collected with the ATLAS detector at the LHC. Events with a photon, an electron or muon, a $b$-tagged jet, and missing transverse momentum are selected. The contribution from events with electrons or hadrons that are misidentified as photons is estimated using data, and the two main background processes with a prompt photon are estimated in control regions. A neural network is used to distinguish the signal and background events, and the data are consistent with the background-only hypothesis. Limits are set on the strength of effective operators that introduce a left- or right-handed flavour-changing $tq\gamma$ coupling with an up-type quark $q$, on the production cross section for FCNC $t\gamma$ production, and on the branching ratio $t\rightarrow \gamma q$. The limits on the branching ratio and on the $t\gamma$ production cross section are the most stringent to date. The resulting limits on the strength of the effective operators are the most stringent limits obtained in searches for events with a $tq\gamma$ vertex.
 
\section*{Acknowledgements}
 
 
We thank CERN for the very successful operation of the LHC, as well as the
support staff from our institutions without whom ATLAS could not be
operated efficiently.
 
We acknowledge the support of ANPCyT, Argentina; YerPhI, Armenia; ARC, Australia; BMWFW and FWF, Austria; ANAS, Azerbaijan; SSTC, Belarus; CNPq and FAPESP, Brazil; NSERC, NRC and CFI, Canada; CERN; CONICYT, Chile; CAS, MOST and NSFC, China; COLCIENCIAS, Colombia; MSMT CR, MPO CR and VSC CR, Czech Republic; DNRF and DNSRC, Denmark; IN2P3-CNRS, CEA-DRF/IRFU, France; SRNSFG, Georgia; BMBF, HGF, and MPG, Germany; GSRT, Greece; RGC, Hong Kong SAR, China; ISF and Benoziyo Center, Israel; INFN, Italy; MEXT and JSPS, Japan; CNRST, Morocco; NWO, Netherlands; RCN, Norway; MNiSW and NCN, Poland; FCT, Portugal; MNE/IFA, Romania; MES of Russia and NRC KI, Russian Federation; JINR; MESTD, Serbia; MSSR, Slovakia; ARRS and MIZ\v{S}, Slovenia; DST/NRF, South Africa; MINECO, Spain; SRC and Wallenberg Foundation, Sweden; SERI, SNSF and Cantons of Bern and Geneva, Switzerland; MOST, Taiwan; TAEK, Turkey; STFC, United Kingdom; DOE and NSF, United States of America. In addition, individual groups and members have received support from BCKDF, CANARIE, CRC and Compute Canada, Canada; COST, ERC, ERDF, Horizon 2020, and Marie Sk{\l}odowska-Curie Actions, European Union; Investissements d' Avenir Labex and Idex, ANR, France; DFG and AvH Foundation, Germany; Herakleitos, Thales and Aristeia programmes co-financed by EU-ESF and the Greek NSRF, Greece; BSF-NSF and GIF, Israel; CERCA Programme Generalitat de Catalunya, Spain; The Royal Society and Leverhulme Trust, United Kingdom.
 
The crucial computing support from all WLCG partners is acknowledged gratefully, in particular from CERN, the ATLAS Tier-1 facilities at TRIUMF (Canada), NDGF (Denmark, Norway, Sweden), CC-IN2P3 (France), KIT/GridKA (Germany), INFN-CNAF (Italy), NL-T1 (Netherlands), PIC (Spain), ASGC (Taiwan), RAL (UK) and BNL (USA), the Tier-2 facilities worldwide and large non-WLCG resource providers. Major contributors of computing resources are listed in Ref.~\cite{ATL-GEN-PUB-2016-002}.
 
 
\printbibliography

\clearpage
 
 
\begin{flushleft}
{\Large The ATLAS Collaboration}

\bigskip

G.~Aad$^\textrm{\scriptsize 101}$,    
B.~Abbott$^\textrm{\scriptsize 128}$,    
D.C.~Abbott$^\textrm{\scriptsize 102}$,    
A.~Abed~Abud$^\textrm{\scriptsize 70a,70b}$,    
K.~Abeling$^\textrm{\scriptsize 53}$,    
D.K.~Abhayasinghe$^\textrm{\scriptsize 93}$,    
S.H.~Abidi$^\textrm{\scriptsize 167}$,    
O.S.~AbouZeid$^\textrm{\scriptsize 40}$,    
N.L.~Abraham$^\textrm{\scriptsize 156}$,    
H.~Abramowicz$^\textrm{\scriptsize 161}$
% ...
\bigskip
\\

% ...
$^{40}$Niels Bohr Institute, University of Copenhagen, Copenhagen; Denmark.\\
% ...
$^{53}$II. Physikalisches Institut, Georg-August-Universit\"{a}t G\"ottingen, G\"ottingen; Germany.\\
% ...
$^{70}$$^{(a)}$INFN Sezione di Pavia;$^{(b)}$Dipartimento di Fisica, Universit\`a di Pavia, Pavia; Italy.\\
% ...
$^{93}$Department of Physics, Royal Holloway University of London, Egham; United Kingdom.\\
% ...
$^{101}$CPPM, Aix-Marseille Universit\'e, CNRS/IN2P3, Marseille; France.\\
$^{102}$Department of Physics, University of Massachusetts, Amherst MA; United States of America.\\
% ...
$^{128}$Homer L. Dodge Department of Physics and Astronomy, University of Oklahoma, Norman OK; United States of America.\\
% ...
$^{156}$Department of Physics and Astronomy, University of Sussex, Brighton; United Kingdom.\\
% ...
$^{161}$Raymond and Beverly Sackler School of Physics and Astronomy, Tel Aviv University, Tel Aviv; Israel.\\
% ...
$^{167}$Department of Physics, University of Toronto, Toronto ON; Canada.\\
% ...
$^{a}$ Also at CERN, Geneva; Switzerland.\\
$^{b}$ Also at CPPM, Aix-Marseille Universit\'e, CNRS/IN2P3, Marseille; France.\\
% ...

\end{flushleft}

 
\end{document}